\documentclass[english,prc,twocolumn,showpacs]{revtex4}
\usepackage[T1]{fontenc}
\usepackage[latin9]{inputenc}
\usepackage{graphicx}
\usepackage{esint}

\providecommand{\tabularnewline}{\\}

\usepackage{babel}

\begin{document}

\title{Relevant energy ranges for astrophysical reaction rates}

\author{Thomas Rauscher}

\affiliation{Department of Physics, University of Basel, Klingelbergstr.\ 82,
4056 Basel, Switzerland}
\begin{abstract}
Effective energy windows (Gamow windows) of astrophysical reaction
rates for (p,$\gamma$), (p,n), (p,$\alpha$), ($\alpha$,$\gamma$),
($\alpha$,n), ($\alpha$,p), (n,$\gamma$), (n,p), and (n,$\alpha$)
on targets with $10\leq Z\leq83$
from proton- to neutron-dripline are calculated using theoretical
cross sections. It is shown that widely used approximation formulae
for the relevant energy ranges are not valid for a large number of reactions
relevant to hydrostatic and explosive nucleosynthesis. The influence
of the energy dependence of the averaged widths on the location of
the Gamow windows is discussed and the results presented in tabular
form.
\end{abstract}


\maketitle

\section{Introduction\label{sec:Introduction}}

Astrophysical reaction rates describe the change in abundances of
nuclei due to nuclear processes in an astrophysical environment, such
as a hot plasma composed of free electrons and atomic nuclei. The
reaction rate per particle pair (or reactivity) is found by folding
interaction cross sections $\sigma$ with the appropriate energy distribution
of the interacting particles in the plasma. For nucleons and nuclei interacting with each other in stars,
the latter is the Maxwell-Boltzmann (MB) distribution, leading to
the definition of the reactivity $\mathcal{R}=F\mathcal{I}$ with
\cite{ili07,rolfs} \begin{eqnarray}
F & = & \sqrt{\frac{8}{\pi\mu}}\left(\frac{1}{kT}\right)^{\frac{3}{2}}\quad,\\
\mathcal{I} & = & \intop_{0}^{\infty}\sigma(E)Ee^{-\frac{E}{kT}}\, dE\quad,\label{eq:integrand}\end{eqnarray}
where $k$ denotes the Boltzmann constant, $T$ the plasma temperature,
and $\mu$ the reduced mass $\mu=M_{1}M_{2}/(M_{1}+M_{2})$. Although
the integration limits run from zero to infinity, the largest contributions
to the integral $\mathcal{I}$ stem from a narrowly confined energy
range, depending on the energy-dependence of the cross sections and
the MB distribution. This relevant energy range has been termed the Gamow
window for charged particle reactions and is important for both nuclear
experimentalists and theoreticians as it defines the energy window
within which the reaction cross sections have to be known.

Due to their importance, simple approximation formulas (see Eqs.\
\ref{eq:e0approx}, \ref{eq:deltaapprox}, \ref{eq:effneutron}, \ref{eq:deltaneutron})
have been derived to estimate
the effective energy windows for reactions (see next sections) and
are frequently used. However, the derivations of these formulae make
implicit assumptions which are not always valid and therefore they
cannot be applied to a number of important cases. For example, it
has been pointed out \cite{ili07,newt07} that resonances below the conventionally
computed Gamow window may contribute significantly to the reaction
rate for narrow-resonance capture of charged particles on light targets.
It will be shown in the following that this can be understood by a
more appropriate treatment of the Gamow window calculation. The applicability
of the approximation will be discussed in more detail and the appropriate
energy windows derived quantitatively for charged-particle induced reactions
(Sec.\ \ref{sec:Charged-particle-reactions}) and for neutral projectiles
(Sec.\ \ref{sec:Reactions-with-neutrons}). Section \ref{sec:Conclusion} concludes with
a discussion of the validity of
the present approach and a brief summary.

\section{Charged-particle reactions\label{sec:Charged-particle-reactions}}

\subsection{The standard approximation of the Gamow window}

The standard approximation of the Gamow window assumes that the energy-dependence
of the cross section $\sigma$ is mainly determined by the projectile's
penetration of the Coulomb barrier. The integral $\mathcal{I}$ can
then be rewritten as \cite{ili07,rolfs} \begin{equation}
\mathcal{I}=\intop_{0}^{\infty}S(E)e^{-\frac{E}{kT}}e^{-2\pi\eta}\, dE\quad,\label{eq:penet}\end{equation}
where $S$ is the astrophysical $S$ factor\begin{equation}
S=\sigma Ee^{2\pi\eta}\label{eq:sfactor}\end{equation}
which is assumed to be only weakly dependent on the energy $E$ for
non-resonant reactions. The second exponential in Eq.\ (\ref{eq:penet})
contains an approximation of the Coulomb penetration through the Sommerfeld parameter
\begin{equation}
\eta=\frac{Z_1 Z_2 e^2}{\hbar}\sqrt{\frac{\mu}{2E}}\quad,
\end{equation}
where $Z_{1}$,
$Z_{2}$ are the charges of projectile and target, respectively, and $\mu$ is the reduced mass. While the first exponential
decreases with increasing energy, this second one increases, leading
to a confined peak of the integrand, the so-called Gamow peak. The
location of the peak $E_{0}$ is shifted to higher energies with respect
to the maximum of the MB distribution at $E_{\mathrm{MB}}=kT$. Assuming
a constant $S$ factor, $E_{0}$ can be determined analytically to
be \cite{ili07,rolfs}\begin{equation}
E_{0}=\left(\frac{\mu}{2}\right)^{\frac{1}{3}}\left(kT\right)^{\frac{2}{3}}\quad.\label{eq:gamowenergy}\end{equation}
The peak is not symmetrical around $E_{0}$ but nevertheless is often
approximated by a Gaussian function\begin{equation}
\mathcal{I}(E)=\mathcal{I}_{\mathrm{max}}e^{-\frac{4\left(E-E_{0}\right)^{2}}{\Delta^{2}}}\quad,\label{eq:gaussian}\end{equation}
where $\mathcal{I}_{\mathrm{max}}=\exp\left(-3E_{0}/(kT)\right)$
is the maximal value of the product of the two exponentials in Eq.\ (\ref{eq:penet})
and $\Delta=4\sqrt{E_{0}kT/3}$ is the 1/e width of the peak. Inserting
the proper numerical factors and units in Eqs.\ (\ref{eq:gamowenergy})
and (\ref{eq:gaussian}) leads to the more practical form \cite{ili07,rolfs,rtk97}\begin{eqnarray}
E_{0} & = & 0.12204\left(\mu_{A}Z_{1}^{2}Z_{2}^{2}T_{9}^{2}\right)^{\frac{1}{3}}\quad,\label{eq:e0approx}\\
\Delta & = & 0.23682\left(\mu_{A}Z_{1}^{2}Z_{2}^{2}T_{9}^{5}\right)^{\frac{1}{6}}\quad.\label{eq:deltaapprox}\end{eqnarray}
Here $E_{0}$ and $\Delta$ are in units of MeV, $T_{9}$ is the plasma
temperature in GK, and $\mu_{A}=A_{1}A_{2}/(A_{1}+A_{2})$ is the
reduced mass number. Equations (\ref{eq:e0approx}) and (\ref{eq:deltaapprox})
are widely used to determine a relevant energy range $E_{0}-\Delta/2\leq E\leq E_{0}+\Delta/2$
within which the nuclear cross sections have to be known. This is
especially important and is often used to design experiments.%
\begin{table}
\caption{\label{tab:truepeaks}Effective energy windows $\widetilde{E}_{\mathrm{hi}}-\widetilde{\Delta}\leq E\leq\widetilde{E}_{\mathrm{hi}}$
for a given plasma temperature $T$. Also given is the energy $\widetilde{E}_{0}$
of the maximum in the reaction rate integrand and its shift $\delta$
relative to the standard formula. The latter is $\delta=\widetilde{E}_{0}-E_{0}$
relative to the location of the Gamow peak $E_{0}$ for charged-particle
induced reactions and $\delta=\widetilde{E}_{0}-E_{\mathrm{MB}}$
relative to the maximum of the MB distribution at $E_{\mathrm{MB}}$
for neutron-induced reactions. This table contains only a few examples.
The full table is available at EPAPS \cite{epaps}.}
\begin{tabular}{ccrrrrr}
\hline \hline
Target & Reaction & \multicolumn{1}{c}{$T$} & \multicolumn{1}{c}{$\widetilde{E}_{\mathrm{hi}}$} & \multicolumn{1}{c}{$\widetilde{\Delta}$} & \multicolumn{1}{c}{$\widetilde{E}_{0}$} & \multicolumn{1}{c}{$\delta$}\tabularnewline
 &  & {[}GK{]} & {[}MeV{]} & {[}MeV{]} & {[}MeV{]} & {[}MeV{]}\tabularnewline
\hline
$^{24}$Mg & ($\alpha$,$\gamma$) & 2.5 & 2.36 & 1.05 & 1.66 & $-1.16$\tabularnewline
$^{27}$Al & (p,$\gamma$) & 3.5 & 1.47 & 1.12 & 0.65 & $-0.89$\tabularnewline
$^{40}$Ca & ($\alpha$,$\gamma$) & 2.0 & 3.62 & 1.39 & 2.85 & $-0.63$\tabularnewline
 &  & 4.0 & 4.66 & 1.97 & 3.56 & $-1.97$\tabularnewline
$^{60}$Fe & (n,$\gamma$) & 5.0 & 1.20 & 1.20 & 0.13 & $-0.30$\tabularnewline
$^{62}$Ni & (n,$\gamma$) & 3.5 & 1.00 & 1.00 & 0.15 & $-0.15$\tabularnewline
$^{106}$Cd & ($\alpha$,$\gamma$) & 3.5 & 10.07 & 3.44 & 8.08 & $-1.17$\tabularnewline
$^{120}$Sn & (n,$\alpha$) & 5.0 & 9.54 & 4.16 & 6.92 & $+6.49$\tabularnewline
$^{144}$Sm & ($\alpha$,$\gamma$) & 3.5 & 11.97 & 3.99 & 9.90 & $-1.10$\tabularnewline
$^{169}$Tm & ($\alpha$,$\gamma$) & 2.0 & 9.20 & 2.94 & 7.61 & $-0.54$\tabularnewline
 &  & 5.0 & 13.20 & 4.27 & 10.22 & $-4.79$\tabularnewline
 \hline
 \hline
\end{tabular}
\end{table}
\begin{figure}
\includegraphics[angle=-90,width=1\columnwidth]{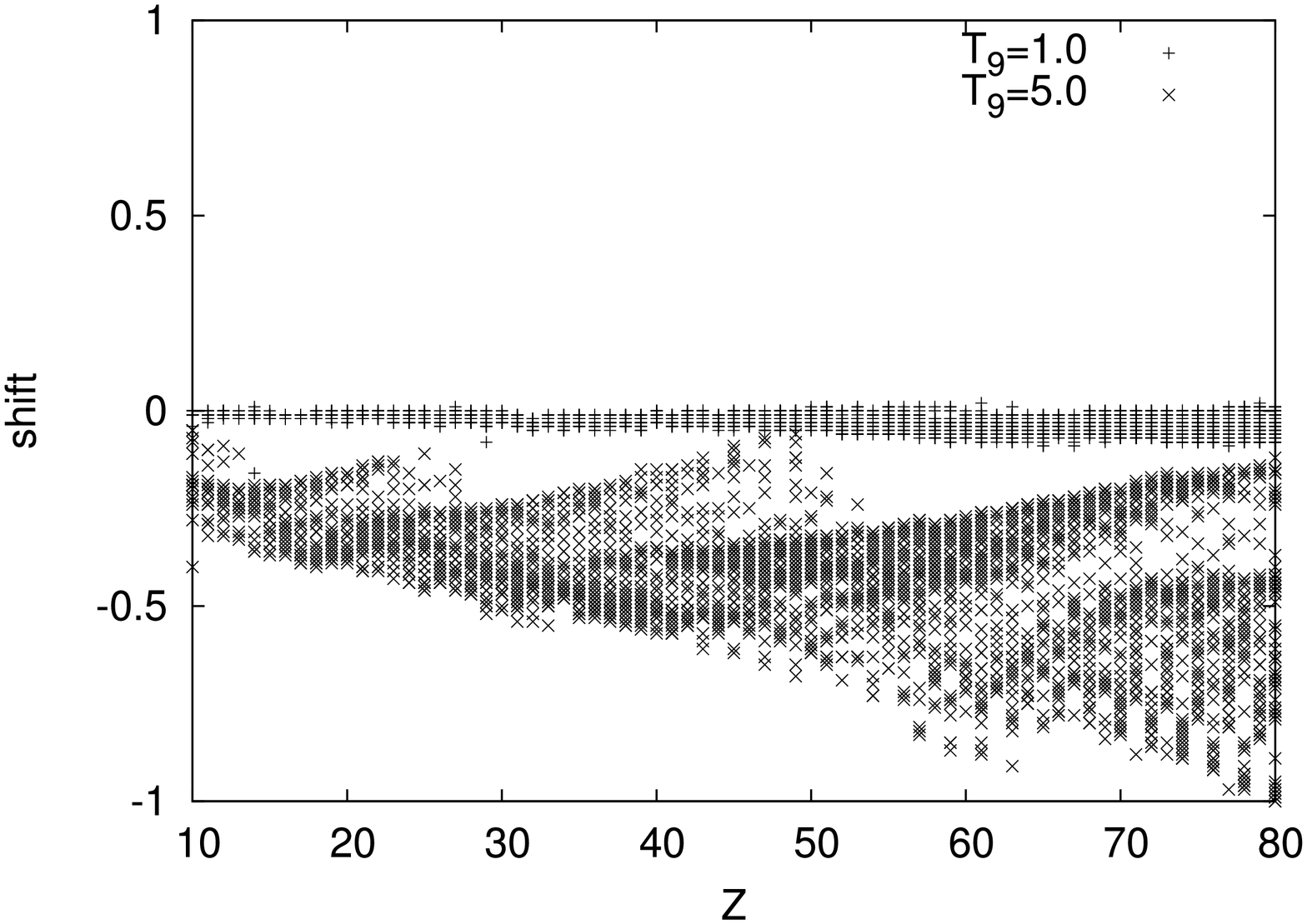}\caption{\label{fig:pn}Shifts $\delta$ (in MeV) of the maximum of the integrand relative
to $E_{0}$ of the Gaussian approximation as a function of the target
charge $Z$ for (p,n) reactions at two temperatures. Almost no shift
is observed at $T_{9}=1.0$ and shifts remain small for $T_{9}=5.0$.}

\end{figure}
\begin{figure}
\includegraphics[angle=-90,width=1\columnwidth]{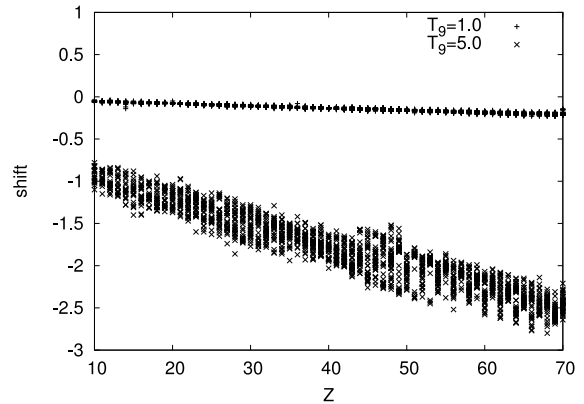}\caption{\label{fig:an}Shifts $\delta$ (in MeV) of the maximum of the integrand relative
to $E_{0}$ of the Gaussian approximation as a function of the target
charge $Z$ for ($\alpha$,n) reactions at two temperatures. Almost
no shift is observed at $T_{9}=1.0$ and shifts reach a few MeV for
$T_{9}=5.0$.}

\end{figure}

\subsection{Criticism of the standard approximation\label{sub:Criticism}}

The derivation of Eq.\ (\ref{eq:gamowenergy}) -- and hence of Eqs.\ (\ref{eq:e0approx}),
(\ref{eq:deltaapprox}) -- implicitly assumes that the energy-dependence
of the cross section $\sigma$ is dominated by the Coulomb barrier
penetration of the projectile. In other words, the energy-dependence
of the entrance channel width dominates. Resonant cross sections can
be written as a sum of Breit-Wigner terms. It can be shown \cite{descrau}
that for a sufficiently high nuclear level density at the compound
nucleus formation energy the sum of overlapping resonances can be
replaced by a sum over averaged widths (or averaged strength functions),
leading to the well-established statistical model of compound reactions
(Hauser-Feshbach model). Thus, both resonant and Hauser-Feshbach cross
sections can be expressed as \cite{descrau}
\begin{equation}
\sigma\propto\sum_{n}(2J_{n}+1)\frac{X_{\mathrm{in}}^{J_{n}}X_{\mathrm{fi}}^{J_{n}}}{X_{\mathrm{tot}}^{J_{n}}}\quad,\label{eq:cs}\end{equation}
with $X$ being either Breit-Wigner widths or averaged Hauser-Feshbach
widths, depending on the context. The width of the entrance channel
is given by $X_{\mathrm{in}}^{J_{n}}$, the one of the exit channel
by $X_{\mathrm{fi}}^{J_{n}}$, and the total width including all possible
emission channels from a given resonance or compound state with spin
$J_{n}$ by $X_{\mathrm{tot}}^{J_{n}}=X_{\mathrm{in}}^{J_{n}}+X_{\mathrm{fi}}^{J_{n}}+\ldots$
Even in the case of the statistical model only few summands in Eq.\ (\ref{eq:cs})
contribute although the sum runs over all values of $J$.

It has become common knowledge that a cross section of the form shown
in Eq.\ (\ref{eq:cs}) is determined by the properties of the smaller
width in the numerator if no other channels than the entrance and
exit channel contribute significantly to $X_{\mathrm{tot}}^{J_{n}}$.
Then $X_{\mathrm{tot}}^{J_{n}}$ cancels with the larger width in
the numerator and the smaller width remains. (The effect is less pronounced
and requires more detailed investigation when other channels are non-negligible
in $X_{\mathrm{tot}}^{J_{n}}$.) In consequence, the energy-dependence
of the cross section will then be governed by the energy dependence
of this smallest $X^{J}$. Only if this happens to be a charged-particle
(averaged) width in the entrance channel, the use of the standard
formula for the Gamow window (Eqs.\ \ref{eq:e0approx}, \ref{eq:deltaapprox})
will be justified. Since $X_{\mathrm{in}}^{J}$ and $X_{\mathrm{fi}}^{J}$
have different energy dependences, it will depend on the specific
energy (weighted by the MB distribution) which of the widths is smaller.
Higher plasma temperature will select an energy range at higher energy.
Therefore, the relative sizes of the $X^{J}$ and thus the energy
dependence of the cross sections in the relevant energy range will
depend on the plasma temperature. This determines the Gamow windows.

It can be seen immediately that the prerequisite for applying the
standard formula may not be met when studying reactions like (p,$\alpha$)
at moderate reaction $Q$ values where one would expect the $\alpha$
width to be smaller than the proton width. Also for charged particle
capture the standard formula will be problematic as long as the $\gamma$
width is smaller than the width in the entrance channel. Although
astrophysically relevant interaction energies are so small that charged
particle widths often become smaller than the radiation widths due
to the Coulomb barrier, reactions in high temperature environments
-- such as explosive nucleosynthesis -- may still exhibit the opposite
relation. Because of the sensitivity of the Gamow window to the relative
sizes of the widths and the complicated dependence of the widths on
the interaction energy, Gamow windows should be extracted from numerical
inspection of the integrand in Eq.\ (\ref{eq:integrand}) and not
from the standard formula which is only applicable for a limited number
of cases. This is performed in the following
sections.%
\begin{figure}
\includegraphics[angle=-90,width=1\columnwidth]{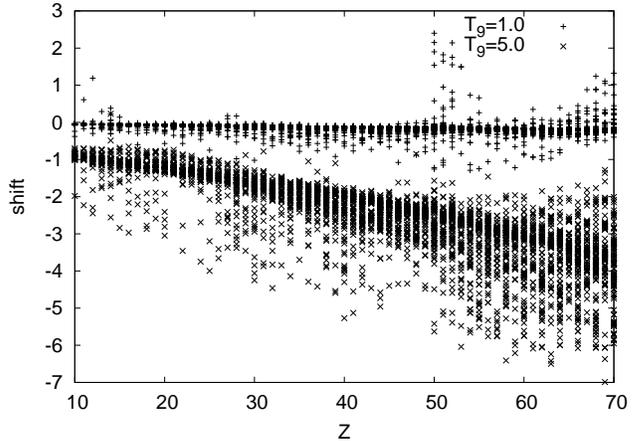}\caption{\label{fig:ag-vs-z}Shifts $\delta$ (in MeV) of the maximum of the integrand
relative to $E_{0}$ of the Gaussian approximation as a function of
the target charge $Z$ for ($\alpha$,$\gamma$) reactions at two
temperatures. Almost no shift is observed at $T_{9}=1.0$ but shifts
become large at $T_{9}=5.0$.}

\end{figure}
\begin{figure}
\includegraphics[angle=-90,width=1\columnwidth]{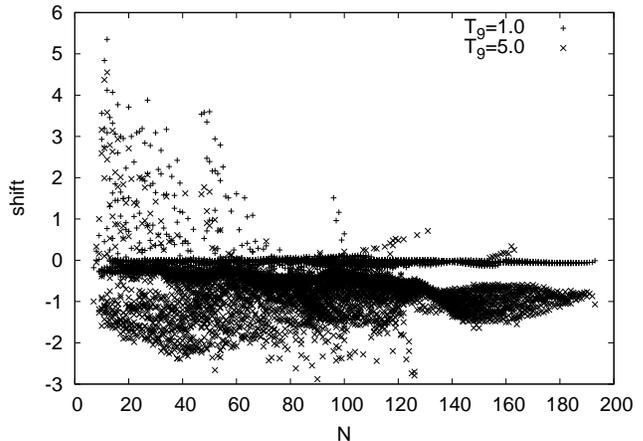}\caption{\label{fig:pg-vs-n}Shifts $\delta$ (in MeV) of the maximum of the integrand
relative to $E_{0}$ of the Gaussian approximation as a function of
the target neutron number $N$ for (p,$\gamma$) reactions at two
temperatures. Almost no shift is observed at $T_{9}=1.0$, except
for proton-rich nuclei with negative reaction $Q$ value. Shifts remain
smaller than for ($\alpha$,$\gamma$) at $T_{9}=5.0$.}

\end{figure}

\subsection{Numerical calculation of Gamow windows}

The function \begin{equation}
\mathcal{F}(E)=\sigma(E)Ee^{-\frac{E}{kT}}\label{eq:integrandfunction}\end{equation}
was computed for all targets and reactions given in \cite{rath01}.
The energies of the maxima $\widetilde{E}_{0}$ and the widths $\widetilde{\Delta}$
of the peaks of $\mathcal{F}$ were determined for each case. A full
table of the results is available as a machine-readable file from
EPAPS \cite{epaps} or at the author's website \cite{nucastro}. With
the exception of captures, only reactions with positive reaction $Q$
value are shown. 
The effective energy window of the reverse reaction
can be found by shifting the energy window by the $Q$ value. For
captures also reactions with $Q<0$ are shown, similar to what is
given in \cite{rath01} because to minimize stellar plasma effects due to thermal
population of the target states it is always preferable to measure captures instead of
photodisintegrations \cite{kiss,raucoulsupp}. Table \ref{tab:truepeaks}, shown here, is
only an abbreviated table to illustrate the kind of information contained
in the full table for a few selected examples. The full table contains
the Gamow windows of 20540 charged-particle reactions for temperatures
$0.5\leq T\leq5.0$ GK, involving targets from proton- to neutron-dripline
and with charge numbers $10\leq Z\leq83$.

It should be noted that the definition of the width $\widetilde{\Delta}$
of the ``true'' Gamow window used here differs from the definition
of the width $\Delta$ as used in Eq.\ (\ref{eq:deltaapprox}). The
width from the standard approximation is the $1/e$ width of the Gaussian
function. However, caution is advised whenever using the energy range
defined in such a manner for, e.g., deriving the energies at which
experimental cross sections are to be determined. For the numerically
derived ``true'' Gamow windows of this work I found that the area
of the peak limited by the $1/e$ width contains only $55-70$\% of
the total contribution to the integral, which would lead to a similar
uncertainty in the reaction rate even when the cross sections within
the window were experimentally completely determined. To obtain a
better measure for the relevant energy range, for each case (of given
reaction on specific target at given temperature) an energy range
was numerically determined contributing 90\% of the total integral
of Eq.\ (\ref{eq:integrand}). This is the width $\widetilde{\Delta}$
quoted in Table \ref{tab:truepeaks} and in the EPAPS table \cite{epaps}.

The energy of the maximum at $\widetilde{E}_{0}$ determines where
the cross sections have the largest weight in the integral. In most
cases, however, the relevant energy window is not symmetric around
this energy. For a more accurate specification of the energy window,
I also give the energy $\widetilde{E}_{\mathrm{hi}}$ of the upper
end of the window in the tables. Thus, the range of relevant energies
is defined as \begin{equation}
\widetilde{E}_{\mathrm{hi}}-\widetilde{\Delta}\leq E\leq\widetilde{E}_{\mathrm{hi}}\quad.\label{eq:defwindow}\end{equation}

In the following I present general observations and discuss selected
cases of interest. For comparison to the standard approximation formula,
I define a shift $\delta=\widetilde{E}_{0}-E_{0}$
relative to the location of the Gamow peak energy $E_{0}$ as given by the
standard approach. In general, at higher temperature and for increasing charge
number the shifts $\left|\delta\right|$ will
be larger. These main effects are
moderated, however, by the involved interplay of different Coulomb
barriers in different channels and different reaction $Q$ values
leading to a more complicated energy dependence of the widths and
thus also of the cross sections.

Since neutrons do not experience a Coulomb barrier and neutron widths
are larger than charged particle widths at low energies in most cases,
the charge and temperature effects can be most clearly seen in (p,n) and
($\alpha$,n) reactions (Figs.$\,$\ref{fig:pn}, \ref{fig:an}).
In both types of reactions there are no or only small shifts at $T_{9}=1.0$
but large shifts at $T_{9}=5.0$. The magnitude of the shifts is generally
larger for the ($\alpha$,n) reactions (note the different scale of
the figures) due to the higher charge of the $\alpha$ particle. The
$\left|\delta\right|$ also increase with increasing target charge
$Z$ in both cases. The shifts are always to smaller energy because
the energy dependence of the neutron widths selects an effective energy
close to $E_{\mathrm{MB}}$ (see Sec. \ref{sec:Reactions-with-neutrons})
which is always below $E_{0}$.%
\begin{figure}
\includegraphics[angle=-90,width=1\columnwidth]{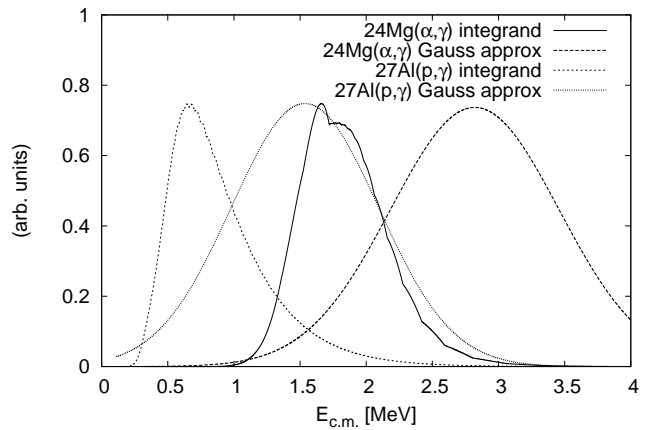}

\caption{Comparison of actual reaction rate integrand $\mathcal{F}$ and Gaussian
approximation of the Gamow window for the reactions $^{24}$Mg($\alpha$,$\gamma$)$^{28}$Si
at $T=2.5$ GK and $^{27}$Al(p,$\gamma$)$^{28}$Si at $T=3.5$ GK.
The integrands and the Gaussians have been arbitrarily scaled to yield
similar maximal values.\label{fig:Comparison-Mg-Al}}

\end{figure}
\begin{figure}
\includegraphics[angle=-90,width=1\columnwidth]{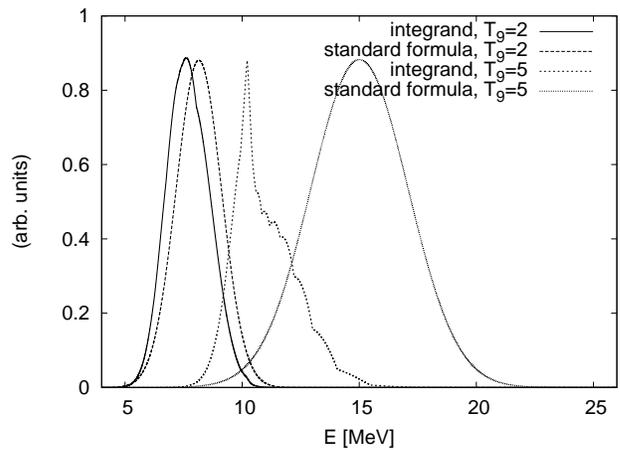}\caption{\label{fig:tm169ag}Comparison of actual reaction rate integrands
$\mathcal{F}$ and Gaussian approximations of the Gamow window for
the reaction $^{169}$Tm($\alpha$,$\gamma$)$^{173}$Lu at $T=2$
and 5 GK. The integrands and the Gaussians have been arbitrarily scaled
to yield similar maximal values. While the shift is small for $T_{9}=2$,
it is about 5 MeV at $T_{9}=5$. Also the asymmetry of the integrand
can be clearly seen at $T_{9}=5$.}

\end{figure}

Charged-particle capture is important in a variety of astrophysical
processes. As mentioned in Sec.$\,$\ref{sub:Criticism}, for capture
reactions the applicability of the standard approximation depends
on the size of the $\gamma$ width $X_{\gamma}^{J}$ relative to the
projectile width $X_{\mathrm{in}}^{J}$. When $X_{\gamma}^{J}\ll X_{\mathrm{in}}^{J}$,
one would assume that there is no Gamow window as the energy dependence
of the $\gamma$ width does not show as strong an increase with increasing
energy as a charged-particle width \cite{ili07}. However, effectively
the Gamow window is shifted only to much lower energy. This can be
understood by the fact that the integration limit in Eq.\ (\ref{eq:integrand})
starts at zero energy and thus will always either include a region
where $X_{\mathrm{in}}^{J}\ll X_{\mathrm{\gamma}}^{J}$ and the Coulomb
penetration is competing with the decay of the MB distribution at
larger energies, or the low-energy region of the MB distribution suppressing
a weakly energy-dependent radiation width. Both cases lead to a peak
in the integrand $\mathcal{F}$ although the ``peak'' may be located
so that it closely approaches zero energy. Figure \ref{fig:ag-vs-z}
shows a similar temperature dependence of the shifts for $\alpha$
captures as for ($\alpha$,n), although the magnitude of the shifts
is larger. These shifts are caused by the fact that at higher $T$
energy regions with $X_{\alpha}^{J}\gg X_{\gamma}^{J}$ receive a
larger weight by the MB distribution. Positive shifts appear for cases
with $Q<0$, simply because the Gamow energy $E_{0}$ derived from
Eq.\ (\ref{eq:e0approx}) is below the ($\alpha$,$\gamma$) threshold
and the actual energy window opens at higher energy. The situation
is similar for (p,$\gamma$) reactions but the temperature dependence
is not as pronounced. The positive shifts occur for proton-rich targets
at the proton dripline, as can be seen in Fig.$\,$\ref{fig:pg-vs-n},
plotting the shifts versus the neutron number $N$.%
\begin{figure}
\includegraphics[angle=-90,width=1\columnwidth]{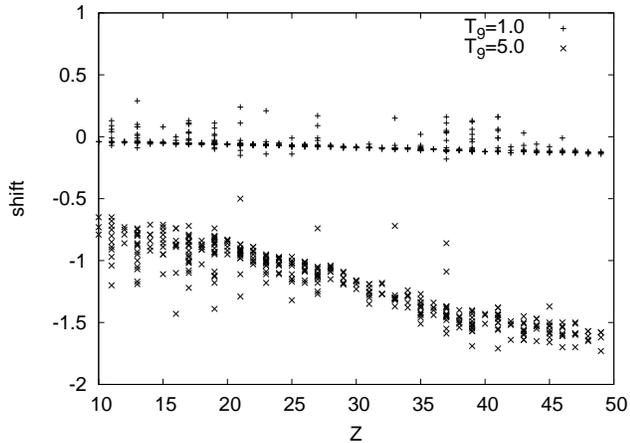}\caption{\label{fig:ap}Shifts $\delta$ (in MeV) of the maximum of the integrand relative
to $E_{0}$ of the Gaussian approximation as a function of the target
charge $Z$ for ($\alpha$,p) reactions at two temperatures. Almost
no shift is observed at $T_{9}=1.0$ and shifts reach a few MeV for
$T_{9}=5.0$.}

\end{figure}
\begin{figure}
\includegraphics[angle=-90,width=1\columnwidth]{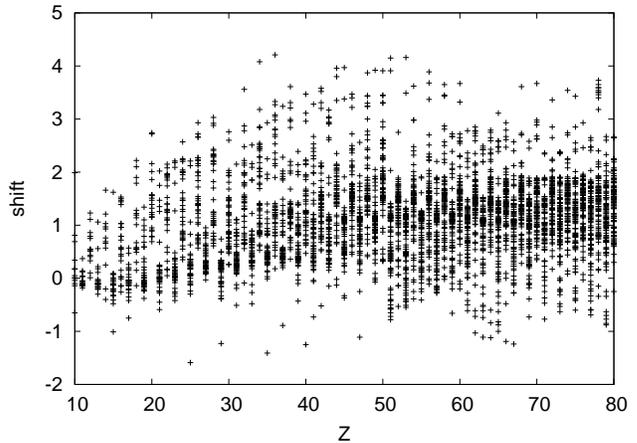}\caption{\label{fig:pa}Shifts $\delta$ (in MeV) of the maximum of the integrand relative
to $E_{0}$ of the Gaussian approximation as a function of the target
charge $Z$ for (p,$\alpha$) reactions at $T=5$ GK. The shifts are
larger as for ($\alpha$,p) reactions and they are positive.}

\end{figure}

The astrophysical importance of capture reactions warrants study for
a few cases in more detail. As already pointed out \cite{ili07,newt07},
it was experimentally found that resonances below the Gamow window,
as defined by the standard approximation formulae, significantly contribute
to the reaction rate for certain capture reactions, e.g., $^{24}$Mg($\alpha$,$\gamma$)$^{28}$Si
and $^{27}$Al(p,$\gamma$)$^{28}$Si. The results for these reactions
are given in Table \ref{tab:truepeaks}. Figure \ref{fig:Comparison-Mg-Al}
shows a comparison of the actual integrands $\mathcal{F}$ and the
Gaussian functions obtained by application of Eqs.\ (\ref{eq:e0approx}),
(\ref{eq:deltaapprox}). This can be directly compared to figure 3.24
of \cite{ili07} where the relative contributions of resonances are
compared to the Gamow window derived from the standard approximation
(for a brief discussion of the relevance of the Gamow window for narrow
resonances, see Sec.$\,$\ref{sec:Conclusion}). The present results
show that the approximation is not valid and the actual Gamow window
is shifted to lower energy in agreement with what was found in \cite{ili07}
but quantifying the relevant energy window. A similar case is $^{40}$Ca($\alpha$,$\gamma$)$^{44}$Ti,
where the effective energy window is also considerably shifted to
lower energy. A plot comparing the actual integrand of the reaction
rate with the Gaussian approximation can be found in \cite{ca40paper}.
A further example is the reaction $^{169}$Tm($\alpha$,$\gamma$)$^{173}$Lu,
shown in Fig.$\,$\ref{fig:tm169ag}. Again, the shift is considerable
at a temperature reached in explosive nucleosynthesis. It is larger
than the shifts for lighter targets because of the larger Coulomb
barrier. The increasing asymmetry of the peak with increasing temperature
can also clearly be seen.

\begin{figure}
\includegraphics[angle=-90,width=1\columnwidth]{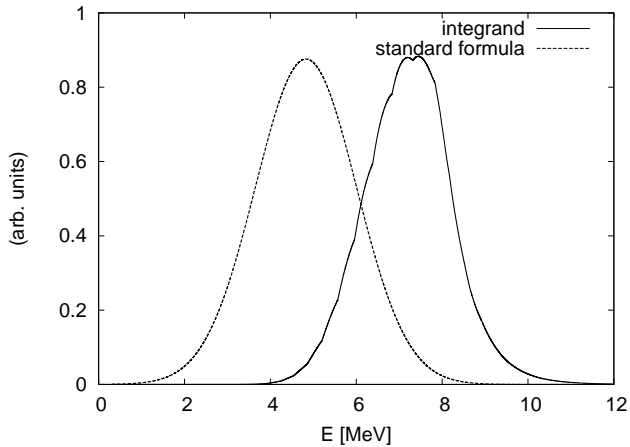}\caption{\label{fig:sn112pa}Comparison of the actual reaction rate integrand
$\mathcal{F}$ and the Gaussian approximation of the Gamow window
for the reaction $^{112}$Sn(p,$\alpha$)$^{109}$In at $T=5$ GK.
The two curves have been arbitrarily scaled to yield similar maximal
values. The maximum of the integrand is shifted by several MeV to
energies higher than the maximum $E_{0}$ of the Gaussian.}

\end{figure}

Finally, we might expect that reactions with charged particles in
both channels show the most complicated dependence on temperature
and charge. However, it is found that the obtained shifts are negligible
at low temperature, which can be explained by the fact that the energy
dependence of the entrance width dominates at low interaction energies.
Figure \ref{fig:ap} shows that at $T_{9}=5$ similar values of the
shifts are obtained in ($\alpha$,p) reactions as in ($\alpha$,n)
reactions, smaller than for ($\alpha$,$\gamma$). The shifts are
also negligible at low temperature for (p,$\alpha$) reactions but
larger and \emph{positive} ($\delta>0)$ shifts are found at high
temperature (see Fig.$\,\ref{fig:pa}$). This is because the proton
widths quickly become larger than the $\alpha$ widths at higher energies
and this leads to a dominance of the $\alpha$ width energy dependence.
Due to the higher Coulomb barrier, the effective energy window is
then found at higher energy as compared to the Gamow window calculated
for protons. This effect is shown in more detail in Fig.$\,$\ref{fig:sn112pa}
for $^{112}$Sn(p,$\alpha$)$^{109}$In at $T_{9}=5.0$.

\section{Reactions with neutrons\label{sec:Reactions-with-neutrons}}

Neutrons are not subject to a Coulomb barrier and therefore a Gamow
peak cannot be defined as in Eq.\ (\ref{eq:penet}). Nevertheless,
an effective energy range can be found because $\sigma(E)$ usually
is a slowly varying function which can be parameterized depending
on the dominant partial wave, i.e., $\sigma\propto1/\sqrt{E}$ for
s-waves, $\sigma\propto\sqrt{E}$ for p-waves, and $\sigma\propto E^{3/2}$
for d-waves. Then the integrand $\mathcal{F}$ defined in Eq.\ (\ref{eq:integrand})
will exhibit a peak mainly determined by the peak of the MB distribution
$E_{\mathrm{MB}}=kT$, only slightly shifted for partial waves $\ell>0$
due to the angular momentum barrier. An often used approximation is
\cite{rtk97,wagoner}\begin{eqnarray}
E_{\mathrm{eff}} & \approx & 0.172T_{9}\left(\ell+\frac{1}{2}\right)\quad,\label{eq:effneutron}\\
\Delta_{\mathrm{eff}} & \approx & 0.194T_{9}\sqrt{\ell+\frac{1}{2}}\quad,\label{eq:deltaneutron}\end{eqnarray}
giving the effective energy window $E_{\mathrm{eff}}\pm\Delta_{\mathrm{eff}}/2$
in MeV for neutrons with energies less than the centrifugal barrier.
This expression is not as handy as the one for the charged-particle
Gamow peak because the dominant partial wave is not known a priori.
Nevertheless, the shifts with partial wave are comparatively small
and the assumption of s-waves is sufficient to define a useful energy
window.%
\begin{figure}
\includegraphics[angle=-90,width=1\columnwidth]{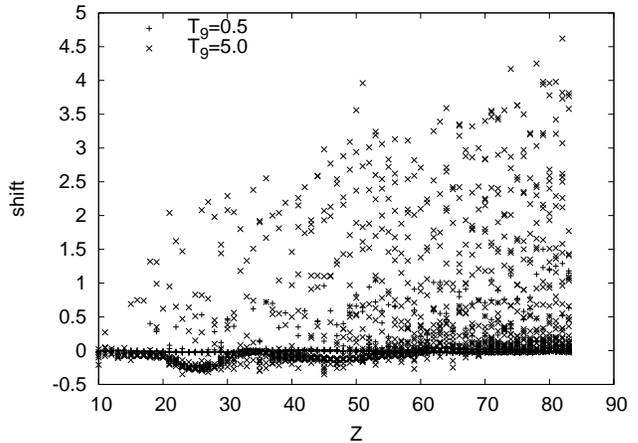}\caption{\label{fig:np}Shifts $\delta$ (in MeV) of the maximum of the integrand relative
to $E_{\mathrm{MB}}$ as a function of the target charge $Z$ for (n,p)
reactions at two temperatures. Almost no shift is observed at $T_{9}=0.5$
but shifts become large at $T_{9}=5.0$.}
\end{figure}
\begin{figure}
\includegraphics[angle=-90,width=1\columnwidth]{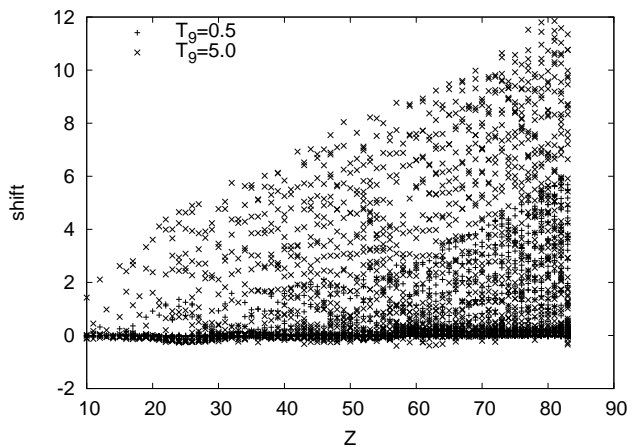}\caption{\label{fig:na}Shifts $\delta$ (in MeV) of the maximum of the integrand relative
to $E_{\mathrm{MB}}$ as a function of the target charge $Z$ for (n,p)
reactions at two temperatures. Almost no shift is observed at $T_{9}=0.5$
but shifts become very large at $T_{9}=5.0$.}
\end{figure}

A straightforward application of Eqs.\ (\ref{eq:effneutron}), (\ref{eq:deltaneutron})
will not be valid in all cases. Just as it was the case with the approximation
for charged projectiles in Sec.\ \ref{sec:Charged-particle-reactions},
the derivation of the formula implicitly assumes that the energy dependence
of the cross section is determined by the neutron projectile. This
is justified for low-energy direct capture or reactions through wings
of broad resonances. In the general resonant case (and the case of
averaging over a large number of resonances as performed in statistical
model calculations), however, the energy dependence of the exit channel
width can become important and strongly modify the effective energy
window.

Similar to the reactions with charged projectiles discussed in Sec.\ \ref{sec:Charged-particle-reactions},
the actual effective energy windows were derived by calculating $\mathcal{F}(E)$
from Eq.\ (\ref{eq:integrandfunction}) and numerically determining
its maximum and the range of the main contributions to the rate integral
(Eq.\ \ref{eq:integrand}). Neutron capture as well as (n,p) and
(n,$\alpha$) reactions are also included in the EPAPS table \cite{epaps}.
Just as for charged particle reactions only reactions with positive
$Q$ value are listed, except for (n,$\gamma$) reactions which also
include a number of endothermic neutron captures at the dripline.
This amounts to a total of 8862 neutron-induced reactions. Table \ref{tab:truepeaks}
shows a few selected examples. The shifts $\delta$ given for neutron-induced
reactions are relative to the peak of the MB distribution as this
defines the location of the peak for neutron s-waves: $\delta=\widetilde{E}_{0}-E_{\mathrm{MB}}$.
The effective energy window is again given by Eq.\ (\ref{eq:defwindow}).

The shifts are zero or small for most neutron captures, except for
those close to the neutron dripline. One would assume that this confirms
the validity of Eqs.\ (\ref{eq:effneutron}), (\ref{eq:deltaneutron})
for (n,$\gamma$) reactions. However, it is rather due to the fact
that the total $\gamma$ widths, being smaller than the neutron widths
in Eq.\ (\ref{eq:cs}), have a comparatively weak energy dependence
and do not move the maximum of the integrand $\mathcal{F}$ from the
one given by the MB distribution. This explains why all of the maxima
$\widetilde{E}_{0}$ are more or less identical to $E_{\mathrm{MB}}$
and thus also $\widetilde{E}_{0}\approx E_{\mathrm{eff}}$, with the
$E_{\mathrm{eff}}$ for $\ell=0$. A different behavior is found only
at low neutron separation energies where either $\gamma$ widths show
a stronger energy dependence or neutron widths become the smallest
widths in the reaction channels. Only in the latter case will different
neutron partial waves be important in Eq.\ (\ref{eq:effneutron}).

Despite the small absolute shifts for neutron captures, it becomes apparent
that the relevant energy window as defined by Eq.\ (\ref{eq:defwindow})
reaches down to the reaction threshold if one desires to determine the
reaction rate with high integration accuracy, even at high plasma
temperature (see also the examples in Table \ref{tab:truepeaks}).

\begin{figure}
\includegraphics[angle=-90,width=1\columnwidth]{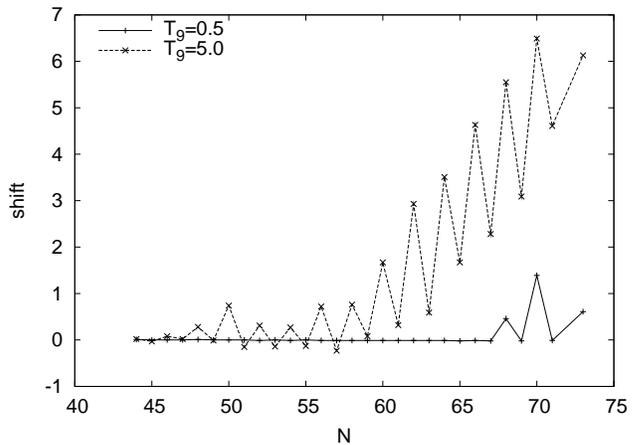}\caption{\label{fig:sn-chain-na}Shifts $\delta$ of the maximum of the integrand
relative to $E_{\mathrm{MB}}$ as function of the target neutron number
$N$ for (n,$\alpha$) reactions on Sn isotopes at two temperatures.
Almost no shift is observed at $T_{9}=0.5$ but shifts become large
at $T_{9}=5.0$ for the neutron-rich isotopes with small reaction
$Q$ value.}

\end{figure}

Larger shifts are found for reactions with charged particles in the
exit channel. Contrary to most cases discussed in Sec.$\,$\ref{sec:Charged-particle-reactions},
the shifts are positive for the majority of the neutron-induced reactions
because the Coulomb barrier penetration in the exit channel moves
the energy window to higher energy. As expected, the shifts are larger
for increased temperature, increased charge of the compound nucleus,
and increased charge of the ejectile. Figures \ref{fig:np} and \ref{fig:na}
illustrate these dependences for (n,p) and (n,$\alpha$) reactions.
For $T_{9}=0.5$ the shifts are close to zero whereas they can reach
several MeV at $T_{9}=5.$ The behavior within an isotopic chain is
shown in Fig.$\,$\ref{fig:sn-chain-na} for (n,$\alpha$) on Sn isotopes.
Again, for low temperature the effective energy window is barely shifted
from the one predicted by the standard formula. With increasing temperature
the shift becomes larger. The shifts for the neutron-rich nuclei are
larger as the energy dependence of the $\alpha$ width becomes stronger
with decreasing $Q$ value.%

\section{Concluding Discussion\label{sec:Conclusion}}

It has to be pointed out that the preceding discussion made implicit
assumptions which have to be scrutinized in any application of the
results. An obvious assumption is that the cross sections used are
correct. The results shown here were obtained with the 'FRDM' set
given in \cite{rath01}. Although transmission coefficients (and thus widths)
are affected by low-lying nuclear states which can be populated in a
reaction, the derived energy \textit{windows} should be robust because the gross energy
dependence of the cross section is the relevant quantity and not its
absolute value. At astrophysical temperatures the MB distribution always
favors energies below the Coulomb barrier. Therefore, the energy dependence of charged-particle
widths is dominated by the Coulomb barrier penetration. Since the energy dependences of widths in
all channels may contribute, the energy windows are also sensitive to the reaction $Q$ values.
Accordingly, updates to masses may affect the conclusions close to the driplines. This also holds
for neutron capture reactions. Although the selection of contributing partial waves strongly
depends on the spectroscopy of a nucleus, the shape of the MB distribution is not strongly modified
when folded with the energy dependence of any relevant partial wave, as explained in Sec.\ \ref{sec:Reactions-with-neutrons}.
Therefore the relevant energy window for neutron capture is defined by the peak of the MB distribution.

Another source of concern may be the use of statistical model Hauser-Feshbach
cross sections for nuclei with low level density close to the driplines,
especially for rates at low plasma temperature. As explained above, the energy windows are mostly determined by the
Coulomb barrier penetration in the different channels or by the MB distribution. This will also hold for direct reactions.
As explained below, the derived effective energy windows remain valid even when narrow
resonances contribute and therefore this is not a limitation of the method.

A final assumption seems to be that the cross sections are smooth
without isolated resonance features. The notion of a single Gamow
peak loses its validity when narrow, isolated resonances are dominating
the reaction rate. In this case, the Gamow window would be fragmented
into several Gamow peaks of different importance, depending on the
resonance strengths. It can be shown, nevertheless, that the notion
of an effective energy window remains valid and that only resonances
within the energy window contribute significantly to the reaction
rate (for details see, e.g., \cite{ili07}). This applies provided the energy windows are derived
as shown above. It does not make a statement about the relative importance
of resonances because this depends on the actual resonance strengths,
not just on the energy dependence of the widths.

Summarizing, a complete numerical study of the effective energy windows
for nuclear reaction rates has been performed for reactions induced
by nucleons and $\alpha$ particles. It has been shown that the actual
energy range of relevant cross sections differs considerably from
the ranges obtained by application of the standard formulae. The origin
of this difference was explained and extensive tables of the actual
energy windows were given. This will be important for further theoretical
improvements of reaction rates as well as for helping to design experiments
to measure cross sections at energies of astrophysical importance.
\begin{acknowledgments}
Discussions with R. D. Hoffman are acknowledged.
This work was supported by the Swiss NSF, grant
200020-122287.
\end{acknowledgments}

\end{document}